\documentclass[letterpaper, 10 pt, conference]{ieeeconf}  

\IEEEoverridecommandlockouts                              
\usepackage{amsmath,amssymb,amsfonts}

\usepackage{algorithm,algorithmic}
\usepackage{graphicx}
\usepackage{textcomp}
\usepackage{color}
\usepackage{lipsum}
\usepackage{epstopdf}
\usepackage{amsopn}
\usepackage{amsthm}
\usepackage{enumerate}
\usepackage{cancel}
\usepackage{mathrsfs}
\usepackage{mathdots}
\usepackage{euscript}
\usepackage{amscd}
\usepackage{cite}
\usepackage{placeins}
\usepackage{tikz}
\usetikzlibrary{snakes,arrows,shapes}
\usepackage{balance}

\theoremstyle{definition}

\theoremstyle{remark}

\newcommand{\bmat}{\begin{bmatrix}}
\newcommand{\emat}{\end{bmatrix}}
\newcommand{\innerprod}[2]{\langle{#1},\,{#2}\rangle}

\DeclareMathOperator{\argmax}{argmax}

\DeclareMathOperator{\E}{{\mathbb E}}
\newcommand{\Rbb}{\mathbb R}

\newcommand{\Zbb}{\mathbb Z}
\newcommand{\Tbb}{\mathbb T}

\newcommand{\yb}{\mathbf  y}
\newcommand{\sbf}{\mathbf s}  

\newcommand{\nb}{\mathbf  n}

\newcommand{\oneb}{\mathbf 1}

\newcommand{\tb}{\mathbf t}
\newcommand{\kb}{\mathbf k}

\newcommand{\Nb}{\mathbf N}

\newcommand{\thetab}{\boldsymbol{\theta}}
\newcommand{\omegab}{\boldsymbol{\omega}}

\newcommand{\zerob}{\boldsymbol{0}}

\renewcommand{\d}{\mathrm{d}}
\newcommand{\p}{\mathrm{p}}
\newcommand{\F}{\mathrm{F}}
\newcommand{\I}{\mathrm{I}}
\renewcommand{\S}{\mathrm{S}}
\renewcommand{\Re}{\mathrm{Re}}

\title{\LARGE \bf Fusion of Sensors Data in Automotive Radar Systems:\\ A Spectral Estimation Approach}

\author{Bin Zhu, Augusto Ferrante, Johan Karlsson, and Mattia Zorzi
\thanks{This work was supported by the SID project ``A Multidimensional and Multivariate Moment Problem Theory for Target Parameter Estimation in Automotive Radars'' (ZORZ\_SID19\_01) funded by the Department of Information Engineering of the University of Padova.}
\thanks{B. Zhu, A. Ferrante, and M. Zorzi  are with the Department of Information Engineering, University of Padova, Via Gradenigo 6/B, 35131 Padova, Italy {\tt\small zhubin@dei.unipd.it}, {\tt\small zorzimat@dei.unipd.it}, {\tt\small augusto@dei.unipd.it}}%
\thanks{J. Karlsson is with Division of Optimization and Systems Theory, Department of Mathematics, KTH Royal Institute of Technology, 10044 Stockholm, Sweden {\tt\small johan.karlsson@math.kth.se}}%
}

\begin{document}

\maketitle
\thispagestyle{empty}
\pagestyle{empty}

\begin{abstract}

To accurately estimate locations and velocities of surrounding targets (cars) is crucial for advanced driver assistance systems based on radar sensors. In this paper we derive methods for fusing data from multiple radar sensors in order to improve the accuracy and robustness of such estimates. First we pose the target estimation problem as 
a multivariate multidimensional spectral estimation problem. The problem is multivariate since each radar sensor gives rise to a measurement channel. Then we investigate
how the use of the cross-spectra affects target estimates. 
We see that the use of the magnitude of the cross-spectrum significantly improves the accuracy of the target estimates, whereas an attempt to compensate the phase lag of the cross-spectrum only gives marginal improvement.
This paper may be viewed as a first step towards applying high-resolution methods that builds on multidimensional multivariate spectral estimation for sensor fusion.

\end{abstract}


\section{INTRODUCTION}

The development of advanced driver assistance systems (ADASs) is one of the key technologies for highly automated automotive systems. Indeed, ADASs can be used in many important tasks such as lane change assistant, forward collision avoidance and adaptive cruise control \cite{eskandarian2012handbook}. The preferred environmental sensing in ADASs is the radar technology. The latter works reliably also in bad lighting conditions and when visibility is reduced due to presence of rain or fog. Radars provide accurate estimates of the target parameters, e.g., range, relative velocity and angle of multiple targets.

State of the art radar systems typically use the chirp sequence modulation principle and a uniform linear array (ULA) of receive antennas. Thus, radar measurements in a coherent processing interval (CPI) is a superposition of $k$ 3-d complex sinusoids where $k$ is the number of targets \cite{gini2012waveform,wintermantel2014radar}. Under reasonable assumptions, such radar measurements can be modeled as a stationary stochastic process whose spectrum is multidimensional and univariate. Moreover, the spectrum is characterized by $k$ peaks (target frequencies) corresponding to the aforementioned sinusoids. Accordingly, an important task is to estimate such a spectrum and, in particular, its peaks and the corresponding frequencies \cite{engels2017advances,engels2014target}. 
Then the target parameters can be recovered from these target frequencies. 

A natural development is towards cars with several automotive radars \cite{lin2015integration,murad2012next}, and thus an important aspect is the integration of multiple radar modules with the aim to improve the target parameter estimation. In the present paper we consider an integrated  system of automotive modules where we have two ULAs of receivers that share one common transmitter. We model the measurements of the two ULAs of receivers as a {\em vector-valued} stochastic process defined in a {\em multidimensional} support.
Hence, its spectrum is a {\em multivariate} (i.e., matrix vauled) complex function defined over a {\em multidimensional} frequency domain. 

We show by simulation evidence that the cross spectrum of the two modules, i.e., the information coming from the dependence between the measurements of the two modules, can be used to improve the estimation of the target frequencies. In doing that, we propose a windowed periodogram to estimate the multidimensional multivariate spectrum from the measurements. 

This work is a first attempt to employ multivariate and multidimensional spectral analysis for high resolution sensor fusion.
Of course a more structured approach in the same vein of  \cite[Section~II]{Georgiou-06} might be employed to cast sensor fusion into a generalized moment problem with an entropic optimization functional. For this kind of problems a very rich stream of research has been produced both in the multidimensional univariate case 
\cite{lang1982multidimensional, georgiou2005solution, ringh2016multidimensional, ringh2018multidimensional} 
and the unidimensional multivariate case \cite{P-F-SIAM-REV,ZORZI201587,avventi_thesis,BLN-03,RFP-09,RFP-10-well-posedness,FPR-08,FMP-12,Z14,Z15,GL-17,Zhu-Baggio-19,georgiou2002spectral,Carli-FPP-11,zhu2018wellposed,zhu2018erratum}. 

The outline of the paper is as follows. In Section \ref{sec:problem}
we formulate the problem. In Section \ref{sec:periodogram} we introduce the multidimensional multivariate windowed periodogram. In Section \ref{sec:sim} simulation experiments are presented. Finally, in Section \ref{sec:concl} we draw the conclusions.

\section{PROBLEM STATEMENT}\label{sec:problem}

\begin{figure*}[!t]
\centering
\includegraphics[width=0.92\textwidth]{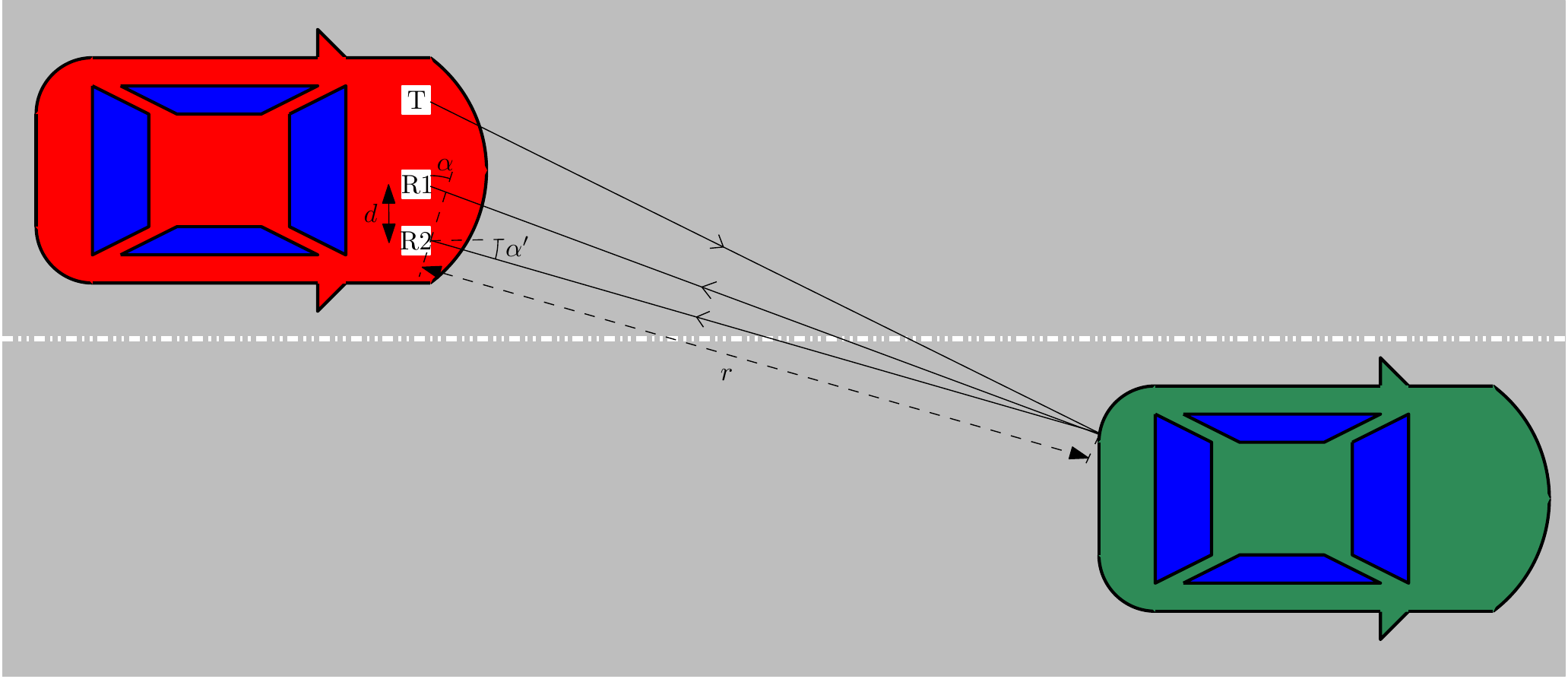}
\caption{An integrated system of automotive modules installed in the red car. T: transmitter, R: ULAs receiver, $d$: distance between the two ULAs, $r$: range of the target, $\alpha$: azimuth angle. Under the far field assumption, we have $\alpha\approx\alpha'$. The green car is the target. }
\label{fig:radar}
\end{figure*}

Consider an automotive radar system that employs coherent linear frequency-modulated (LFM) pulse train signals and uses a 
ULA of receive antennas for the measurement.
Assume, for simplicity, that only one target is present in the field of view.
According to \cite{engels2014target,engels2017advances}, after down-mixing, filtering, and sampling, the (scalar) measurement of a ULA in a
CPI is modeled as a $3$-d complex sinusoidal signal
\begin{equation}\label{y_measurement}
y(\tb) =  a \, e^{ i \left( \innerprod{\thetab}{\tb} +\varphi \right)} + w(\tb).
\end{equation}
The meaning of each variable will be explained next.
The index $\tb$ takes values in the set 
\begin{equation*}
\Zbb_\Nb^3:=\left\{(t_1,t_2,t_3)\in\Zbb^3 \, :\, 0 \leq t_j \leq N_j-1, j=1,2,3 \right\},
\end{equation*}
where $N_1$ denotes the number of samples per pulse, $N_2$ the number of pulses, $N_3$ the number of (receive) antennas, and $\Nb:=[N_1,N_2,N_3]$ defines the size of the data array. The scalar $a$ is a real amplitude. 
The variable $\varphi$ is an initial phase angle which is assumed to be a random variable uniformly distributed in $[-\pi,\pi]$ (cf.~\cite[Section~4.1]{stoica2005spectral}).
The process $w$ is a zero-mean circular complex white noise independent of $\varphi$. 
The real vector $\thetab=(\theta_1,\theta_2,\theta_3)$ contains three \emph{unknown} normalized angular target frequencies so that $\thetab\in\Tbb^3:=[-\pi,\pi]^3$ and 
$\innerprod{\tb}{\thetab}:=t_1\theta+t_2\theta_2+t_3\theta_3$ is the usual inner product in $\Rbb^3$. Moreover, the components $\theta_j$ $(j=1,2,3)$ are related to the range $r$, the (radial) relative velocity $v$, and the azimuth angle $\alpha$ of the target via
\begin{equation}
\label{linthetaeq}
\begin{split}
\theta_1 & = c_1 r-\pi \\
\theta_2 & = c_2 v \\
\theta_3 & = c_3 \sin \alpha \\
\end{split}
\end{equation}
where $c_j$ $(j=1,2,3)$ are known positive constants which may be computed explicitly from parameters of the radar system \cite[Section~16.4]{engels2014target}.
More specifically, these formulas depend on the waveform, the array geometry of the radar and on the sampling rate of the signal processing unit. 
Notice that we follow the convention in which the azimuth angle $\alpha$ ranges from $-\pi/2$ to $\pi/2$ (see Fig.~\ref{fig:radar}).
It is easy to see that the parameter vector $(r,v,\alpha)$ can be readily recovered from the frequency vector $\thetab$. 
In addition, the target range and velocity are assumed to belong to given closed intervals, namely
$r \in[0,r_{\max}]$, $v \in[-v_{\max},v_{\max}]$ where $r_{\max}=2\pi/c_1$ and $v_{\max}=\pi/c_2$. It is reported in \cite{rohling2012continuous} that under their radar implementation $r_{\max}=200$ m and $v_{\max}=250$ km/h. We also assume that the velocity is fixed in one CPI (which is at the scale of 20 ms).

The target parameter estimation problem consists in estimating the unknown target frequencies $\thetab$ from the sinusoid-in-noise measurements generated according to model \eqref{y_measurement}; indeed an estimate of $(r,v,\alpha)$ is straightforwardly given through (\ref{linthetaeq}). 
Such a frequency estimation problem has been well studied in the literature (see, e.g., \cite[Chapter~4]{stoica2005spectral}).
Through elementary calculations, one gets
\begin{equation}
\sigma_{\kb}:= \E y(\tb+\kb) y(\tb)^* = a^2 e^{i\innerprod{\thetab}{\kb}} + \tilde{\sigma}^2 \delta_{\kb,\zerob},
\end{equation}
where $(\cdot)^*$ denotes complex conjugate (transpose), $\tilde{\sigma}^2$ is the noise variance, and $\delta_{\kb,\zerob}$ is the Kronecker delta function. Taking the Fourier transform, the multidimensional-univariate  spectrum of the signal is
\begin{equation}
\Phi(\omegab) = 2\pi a^2 \delta(\omegab-\thetab) + \tilde{\sigma}^2,
\end{equation}
where $\delta(\cdot)$ here is the Dirac delta measure. If an estimate $\hat{\Phi}$ of the spectrum is available, an estimator of $\thetab$ is given by \cite{engels2014target} 
\begin{equation}
\hat{\thetab}:= \underset{\omegab\in\Tbb^3}{\argmax} \, \hat{\Phi}(\omegab).
\end{equation}

In what follows, we consider an integrated system of automotive modules where we have two ULAs of receivers that share one common transmitter. The ULAs are placed in the same line at a distance $d$ {(see Fig.~\ref{fig:radar})}, say a few decimeters. Under the \emph{far field} assumption which is common in this kind of setup, the target parameters $(r,v,\alpha)$ sensed by the two ULAs essentially differ only in the range dimension by $d \sin\alpha$. 
Then the measurement equation becomes
\begin{equation}\label{y_j_measurement}
\begin{split}
y_1(\tb) & = a \, e^{ i \left( \innerprod{\thetab}{\tb} +\varphi \right)} + w_1(\tb) \\
y_2(\tb) & = a \, e^{ i \left( \innerprod{\thetab}{\tb} + M\theta_3 +\varphi \right)} + w_2(\tb) \,.
\end{split}
\end{equation}
Here $M=d/\Delta s$ where $\Delta s$ is the distance between two adjacent antennas in the ULA, and the phase shift $M\theta_3$ represent the phase shift between the measurements of the two ULAs due to the distance $d$. 
The noises in different channels are assumed to be uncorrelated with the same variance $\tilde{\sigma}^2$. Set $\yb(\tb):=[\,y_1(\tb)\,y_2(\tb)\,]^\top$, then we have
\begin{equation}
\Sigma_{\kb}:= \E \yb(\tb+\kb) \yb(\tb)^* = a^2 e^{i\innerprod{\thetab}{\kb}} R + \tilde{\sigma}^2 \delta_{\kb,\zerob} I_2,
\end{equation}
where the matrix
\begin{equation}\label{mat_R}
R=\bmat
1&e^{-iM\theta_3}\\
e^{iM\theta_3}&1
\emat.
\end{equation}
The multidimensional-multivariate spectrum of $\yb$ is
\begin{equation}\label{Phi_ideal}
\Phi(\omegab) = 2\pi a^2 \delta(\omegab-\thetab) R + \tilde{\sigma}^2 I_2.
\end{equation}
Let $\hat \Phi$ be an estimate of $\Phi$.
We need a post-processing method to integrate the information from the two ULAs in order to obtain an estimator of $\thetab$.
The most straightforward way to use the information is to treat the two signals $\yb_1$ and $\yb_2$ independently. 
Then we estimate the two multidimensional-univariate spectra $\hat{\Phi}_{11}(\omegab)$ of $\yb_1$ and 
$\hat{\Phi}_{22}(\omegab)$ of $\yb_2$ separately.
Finally, we compute the estimate as
\begin{equation}\label{theta_est_m1}
\hat{\thetab}_\I:= \underset{\omegab\in\Tbb^3}{\argmax} \, |\hat{\Phi}_{11}(\omegab)|^2 + |\hat{\Phi}_{22}(\omegab)|^2,
\end{equation}
where the subscript $_\I$ of the estimate $\hat{\thetab}$ stands for ``independent''.
This is essentially a univariate philosophy as it merges the estimate of two univariate spectra.

We propose to use a genuinely multivariate approach where the cross-spectrum $\hat \Phi_{12}$ is taken into account and plays a crucial role.
We will test the performances of our approach against  the estimate (\ref{theta_est_m1}) and show that these performances can be 
improved by taking $\hat \Phi_{12}$ into account in a wise manner.
The first method of this multivariate approach considers the estimate
\begin{equation}
\label{theta_est_m3}
\hat{\thetab}_\S\!:= \underset{\omegab\in\Tbb^3}{\argmax} \,|\hat{\Phi}_{11}(\omegab)|^2 + 2\left[\Re(e^{iM\omega_3} \hat{\Phi}_{12}(\omegab))\right]^2 \!+ |\hat{\Phi}_{22}(\omegab)|^2,
\end{equation}
where the subscript $_\S$ of the estimate $\hat{\thetab}$ stands for ``shifted''. In fact, the weight $e^{iM\omega_3}$ for the cross spectrum $\hat{\Phi}_{12}(\omegab)$ was designed to cancel out the phase shift caused by the third component of the (true) target frequency in $\thetab$, namely the $(1,2)$ element of the matrix $R$ in \eqref{mat_R}.
This seems to be a very natural method for taking $\hat \Phi_{12}$ into account. Simulation evidence, however, shows that the improvement of performances with respect to (\ref{theta_est_m1}) is modest (see Section~\ref{sec:sim}).
A better approach appears to be the following:
\begin{equation}\label{theta_est_m2}
\hat{\thetab}_\F:= \underset{\omegab\in\Tbb^3}{\argmax} \, \|\hat{\Phi}(\omegab)\|_{\F}^2,
\end{equation}
where the subscript $_\F$ of the estimate $\hat{\thetab}$ stands for ``Frobenius'' as
$\|\hat{\Phi}(\omegab)\|_{\F}^2:=|\hat{\Phi}_{11}(\omegab)|^2+|\hat{\Phi}_{22}(\omegab)|^2+2\,|\hat{\Phi}_{12}(\omegab)|^2$ is the Frobenius norm.

The above reasoning can be generalized to the case of $n$ targets in a straightforward manner. The measurement equation in that case becomes
\begin{equation}\label{y_meas_n}
\begin{split}
y_1(\tb) & = \sum_{k=1}^{n} a_k e^{ i \left( \innerprod{\thetab^{(k)}}{\tb} +\varphi_k \right)} + w_1(\tb) \\
y_2(\tb) & = \sum_{k=1}^{n} a_k e^{ i \left( \innerprod{\thetab^{(k)}}{\tb} + M\theta^{(k)}_3 +\varphi_k \right)} + w_2(\tb),
\end{split}
\end{equation}
in which the random phases $\{\varphi_k\}$ are assumed to be independent. The spectrum of the signal is a sum of $n$ Dirac deltas with masses concentrated at the target frequencies $\{\thetab^{(k)}\}$ plus the noise spectrum. Then, the frequency estimates can be obtained as the first $n$ peaks of some suitable objective function (e.g., those reported previously) designed from the estimated spectrum. Another option is to use methods such as orthogonal matching pursuit (see, e.g., \cite{bruckstein2009sparse}) or RELAX \cite{li1996efficient} for the post processing step by using the relation \eqref{Phi_ideal}. The remaining part of the problem is how to obtain a good estimate of the multidimensional-multivariate spectrum from a finite number of observations. Note that even though the model is derived for two radar sensors it is straightforward to extend the model to an arbitrary number of aligned sensors.

\section{THE WINDOWED PERIODOGRAM}\label{sec:periodogram}

Suppose that we have a finite realization of the random field $\yb(\tb)$ given by \eqref{y_j_measurement} with $\tb\in\Zbb_\Nb^3$. Define the finite Fourier transform
\begin{equation}\label{y_sample_FT}
\hat{\yb}_{\Nb}(\omegab) := \sum_{\tb\in\Zbb^3_{\Nb}} \yb(\tb) e^{-i\innerprod{\tb}{\omegab}}.
\end{equation}
Then the (unwindowed) \emph{periodogram} is defined as
\begin{equation}\label{Phi_periodgram}
\hat{\Phi}_{\p}(\omegab) := \frac{1}{|\Nb|} \hat{\yb}_{\Nb}(\omegab) \hat{\yb}_{\Nb}(\omegab)^*
\end{equation}
where $|\Nb|:=N_1 N_2 N_3$ (cf. \cite{brockwell1991time}). This is a function on $\Tbb^3$ whose values are Hermitian positive semidefinite matrices of rank one. In practice, $\hat{\Phi}_{\p}$ is an asymptotically unbiased estimator of $\Phi$ due to the relation 
\begin{equation}\label{Phi_def2}
\Phi(\omegab):=\sum_{\kb\in\Zbb^3} \Sigma_{\kb}e^{-i\innerprod{\kb}{\omegab}} = \lim_{\min(\Nb)\to\infty} \E \hat{\Phi}_{\p}(\omegab)
\end{equation}
which holds under a mild assumption on the decay rate of the covariances. The more precise statement is given in \cite[Section~V]{ZFKZ2019M2}. 

It is well-known in the scalar unidimensional case (i.e., a scalar process with a single ``time'' index) that the periodogram corresponds to the correlogram using the \emph{standard biased covariance estimator} for $\Sigma_\kb$ (see e.g., \cite[Chapter~2]{stoica2005spectral}). The same holds in our multivariate and multidimensional case. To see this, we need to first introduce the set
\begin{equation}\label{Zd_2K}
\Zbb^3_{2\Nb-\oneb} := \{ \kb\in\Zbb^3 \,:\, -N_j+1 \le k_j \le N_j-1,\, j=1,2,3 \}
\end{equation}
for the covariance lags.
Then after a change of the summation index, we have
\begin{equation} 
\begin{split}
\hat{\Phi}_{\p}(\omegab) & = \frac{1}{|\Nb|} \sum_{\tb\in\Zbb^3_{\Nb}} \sum_{\sbf\in\Zbb^3_{\Nb}} \yb(\tb) \yb(\sbf)^* e^{-i\innerprod{\tb-\sbf}{\omegab}} \\
 & = \frac{1}{|\Nb|} \sum_{\kb\in\Zbb^3_{2\Nb-\oneb}} \sum_{\sbf\in\Xi_{\Nb,\kb}} \yb(\sbf+\kb) \yb(\sbf)^* e^{-i\innerprod{\kb}{\omegab}},
\end{split}
\end{equation}
where each component of the index $\sbf$ in the second line must satisfy
\begin{equation}\label{s_j_inequal}
\left\{ \begin{array}{ll}
0 \le s_j \le N_j-1-k_j & \textrm{if } k_j \ge 0 \\
-k_j \le s_j \le N_j-1 & \textrm{if } k_j<0 \\
\end{array} \right. ,
\end{equation}
and hence the set
\begin{equation}
\Xi_{\Nb,\kb}:=\{\sbf\in\Zbb^3 \,:\, s_j \textrm{ satisfies \eqref{s_j_inequal} for } j=1,2,3\}.
\end{equation}
For $\kb\in\Zbb^3_{2\Nb-\oneb}$, take the covariance estimate to be
\begin{equation}\label{cov_est}
\hat{\Sigma}_{\kb}:=\frac{1}{|\Nb|}\sum_{\sbf\in\Xi_{\Nb,\kb}} \yb(\sbf+\kb) \yb(\sbf)^*,
\end{equation}
and we have
\begin{equation}\label{Phi_correlogram}
\hat{\Phi}_{\p}(\omegab)= \sum_{\kb\in\Zbb^3_{2\Nb-\oneb}} \hat{\Sigma}_{\kb} e^{-i\innerprod{\kb}{\omegab}}.
\end{equation}

From the above relation, one immediately sees that the periodogram is not a good estimator of the spectrum because the sample covariances with large lags (which are very noisy) enter $\hat{\Phi}_\p$ as if they were as reliable as $\Sigma_{\zerob}$. Moreover, by the definition \eqref{Phi_periodgram}, we know that the periodogram is always singular for any $\omegab\in\Tbb^3$. As a result, in our $2\times2$ scenario, we have $|\Phi_{12}(\omegab)|^2=\Phi_{11}(\omegab)\Phi_{22}(\omegab)$, which means that the cross spectrum does not give much extra information since its modulus is completely determined by the diagonal terms.

To handle these issues concerning the periodogram, we resort to the windowing technique, also know as the Blackman-Tukey method. More precisely, we first fix a vector of positive integers $\nb:=[n_1,n_2,n_3]$ which are small elementwise compared to $\Nb$. Then we choose the index set
\begin{equation}
\Lambda:=\{\kb\in\Zbb^3\,:\,|k_j|\leq n_j,\ j=1,2,3\}
\end{equation}
as the domain of the real-valued window function $w(\cdot)$,
and discard those covariance estimates \eqref{cov_est} with indices $\kb$ outside the set $\Lambda$. The resulting spectrum estimator is
\begin{equation}\label{Phi_corr_windowed}
\hat{\Phi}(\omegab)= \sum_{\kb\in\Lambda} w(\kb) \hat{\Sigma}_{\kb} e^{-i\innerprod{\kb}{\omegab}}.
\end{equation}

\subsection{Some Computational Details}
In order to obtain the windowed periodogram \eqref{Phi_corr_windowed}, we need to compute the covariance estimates \eqref{cov_est}, at least for the indices  $\kb\in\Lambda$. Under our multidimensional-multivariate setup, directly computing those quantities can be messy and asymptotically inefficient. An alternative implied by the relation \eqref{Phi_correlogram} is to first form the unwindowed periodogram using \eqref{y_sample_FT} and \eqref{Phi_periodgram}, and then obtain $\{\hat{\Sigma}_\kb\}$ with the inverse FFT. Notice that we have to respect the signal processing convention of the DFT which is followed by implementations of the FFT routine. More precisely, if we relabel the covariance sequence by letting $\hat{\Sigma}_\kb=X_{\kb+\Nb-\oneb}$, then it is easy to verify that
\begin{equation*}
\sum_{\kb=\zerob}^{2(\Nb-\oneb)} X_\kb e^{-i\innerprod{\kb}{\omegab}} = e^{-i\innerprod{\Nb-\oneb}{\omegab}} \sum_{\kb\in\Zbb^3_{2\Nb-\oneb}} \hat{\Sigma}_\kb e^{-i\innerprod{\kb}{\omegab}}.
\end{equation*}
This means that the steps to get the covariance estimates in \eqref{cov_est} given the data array $\yb(\tb)$, $\tb\in\Zbb_\Nb^3$ are the following three:
\begin{enumerate}
\item Perform the FFT of size $2\Nb-\oneb$ to $\yb(\tb)$ with zero padding;
\item Compute the periodogram $\hat{\Phi}_{\p}(\omegab)$ by \eqref{Phi_periodgram};
\item Perform the inverse FFT of $e^{-i\innerprod{\Nb-\oneb}{\omegab}} \hat{\Phi}_{\p}(\omegab)$. 
\end{enumerate}

Similarly, the windowed periodogram \eqref{Phi_corr_windowed} is obtained by performing FFT of size $\Nb$ to the covariance sequence $\{w(\kb)\hat{\Sigma}_\kb\}_{\kb\in\Lambda}$ and multiplying the result by $e^{i\innerprod{\nb}{\omegab}}$.

In the case of $m$ measurement channels, the ``independent'' processing \eqref{theta_est_m1} needs $m$ auto-spectra. In contrast, the other two schemes \eqref{theta_est_m3} and \eqref{theta_est_m2} require also the cross spectra. Clearly, the time complexity for computing the complete matricial spectrum grows roughly at the scale of $m^2$.

Recall at last that after the windowing operation, the spectrum \eqref{Phi_corr_windowed} is not necessarily positive definite over the frequency domain. Such positivity is checked on the $\Nb$-grid of $\Tbb^3$.

\section{SIMULATION RESULTS}\label{sec:sim}

	\begin{figure}[!t]
		\centering
		\includegraphics[width=0.5\textwidth]{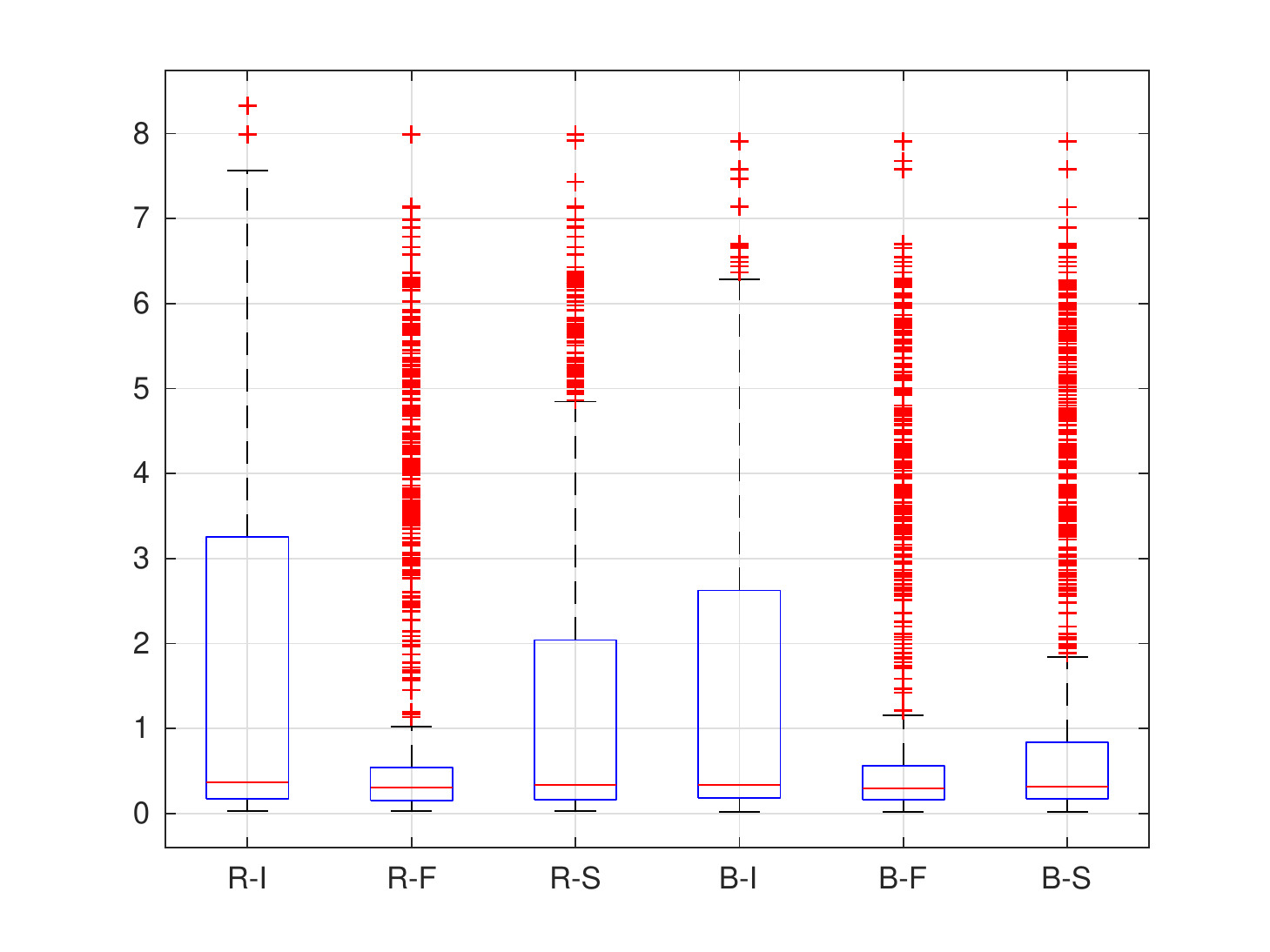}
		\caption{Frequency estimation error when $\Nb=[\,40\;40\;7\,]$.}
		\label{fig:res1}
	\end{figure}
	\begin{figure}[!t]
		\centering
		\includegraphics[width=0.5\textwidth]{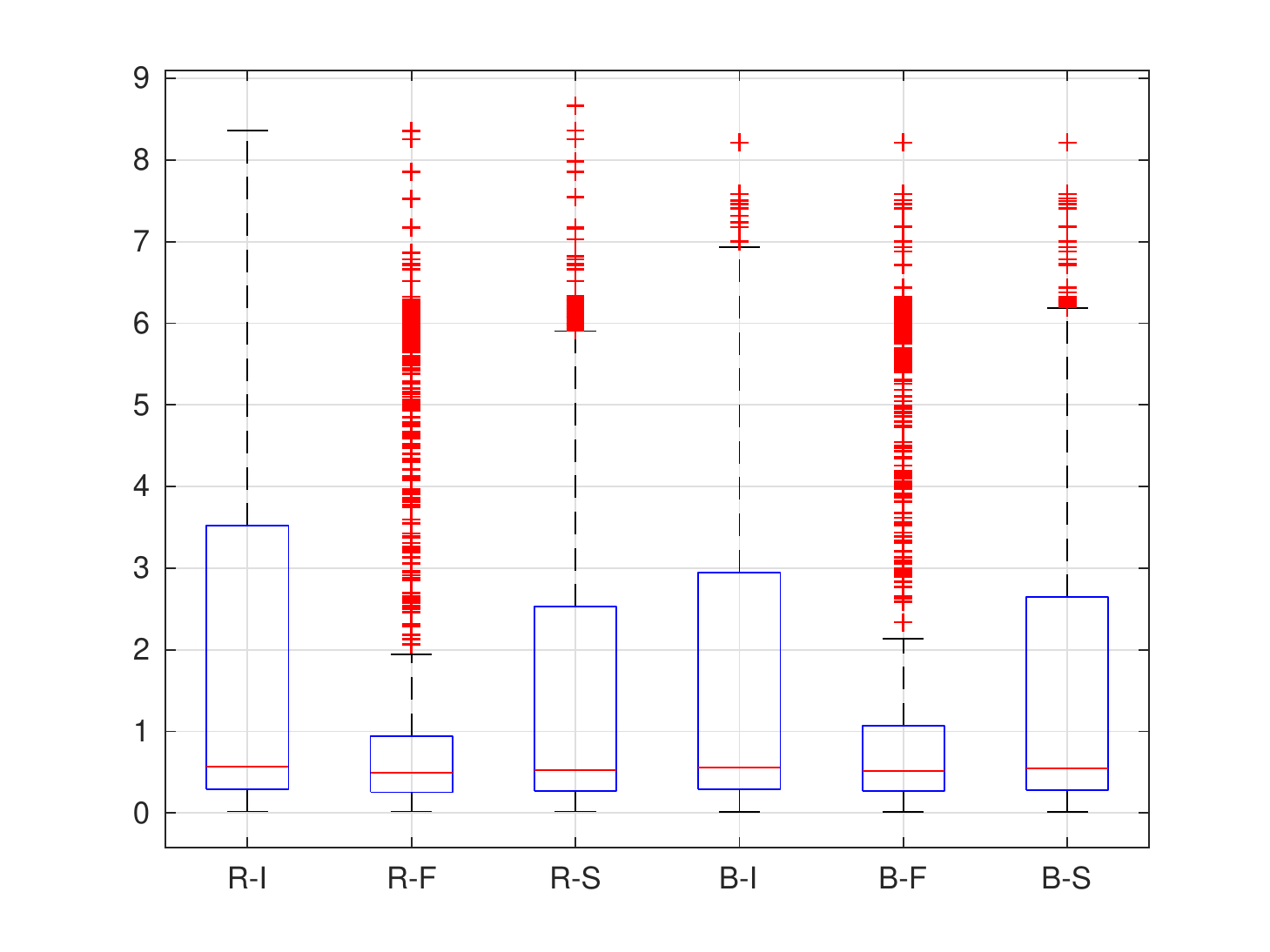}
		\caption{Frequency estimation error when $\Nb=[\,60\;60\;4\,]$.}
		\label{fig:res2}
	\end{figure}
\begin{figure}[!t]
		\centering
		\includegraphics[width=0.5\textwidth]{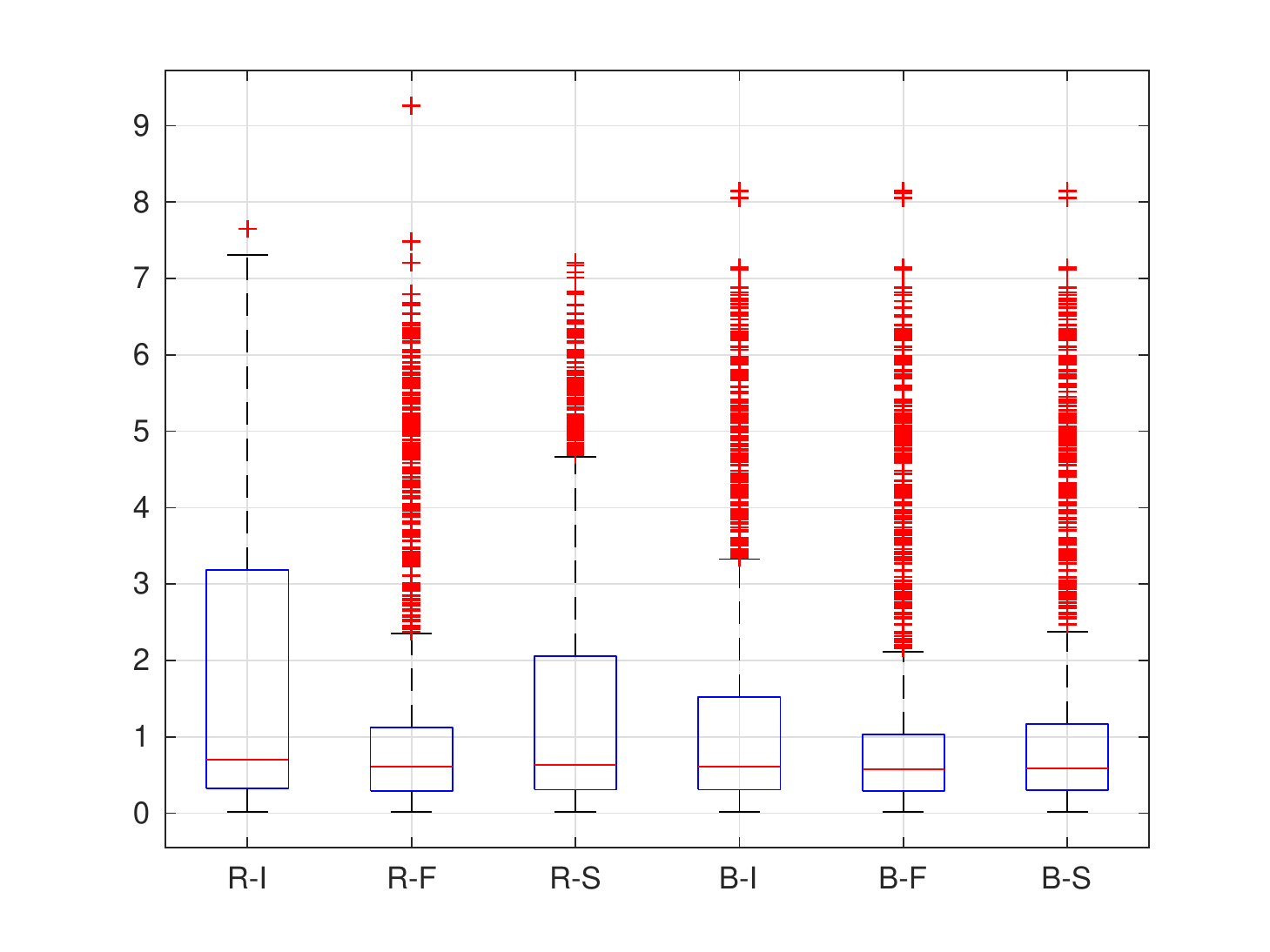}
		\caption{Frequency estimation error when $\Nb=[\,70\;70\;3	\,]$.}
		\label{fig:res3}
	\end{figure}

We perform Monte Carlo simulations, each of which contains $1000$ repeated trials. In each trial, every component of the target frequencies vector $\thetab$ is drawn from the uniform distribution in $[-\pi,\pi]$. 
The signal amplitude is fixed as $a=1$. The measurement noises are zero-mean Gaussian with a standard deviation $\tilde{\sigma}=20$ (low-quality sensor). The integer that appears in the second measurement channel for the phase shift is set as $M=20$.
After choosing the size $\Nb$ of the data array, the vector measurements $\yb(\tb)$ are generated
according to the model \eqref{y_j_measurement}. 
Then the windowed periodogram \eqref{Phi_corr_windowed} is computed. More precisely, the following two window functions have been implemented to smooth the periodogram:
\begin{enumerate}
\item the rectangular window, that is $w(\kb)=1$ for all $\kb\in\Lambda$;
\item the Bartlett window, that is $$w(\kb)=w_1(k_1)w_2(k_2)w_3(k_3)$$ where for $j=1,2,3$,
\begin{equation}
w_j(k_j)=\frac{n_j+1-|k_j|}{n_j+1},\quad k_j=-n_j,\dots,n_j.
\end{equation}
\end{enumerate}
The window widths are fixed as $\nb=[\,8 \; 8 \; 2\,]$ for the rectangular window and $\nb=[\,12\;12\;3\,]$ for the Bartlett window. {These windows have been chosen empirically in such a way that the single channel periodograms exhibit good performances.} Then, the target frequencies vector $\hat{\thetab}$ is obtained by each of the three post-processing methods described in Section~\ref{sec:problem}. Next we consider the following estimators for $\thetab$: R-I, R-F, R-S, B-I, B-F and B-S. R and B means that the windowed periodogram has been computed 
using the rectangular and the Bartlett window, respectively. The letters I, S, and F correspond to the methods for searching the peak in the windowed periodogram using (\ref{theta_est_m1}), (\ref{theta_est_m3}) and (\ref{theta_est_m2}), respectively. The performances of such estimators are measured by the norm of the error, namely $\|\hat{\thetab}-\thetab\|$. These errors in each trial are collected and visualized using the boxplot. According to the Matlab documentation, on each box, the central mark indicates the median, and the bottom and top edges of the box indicate the $25$th and $75$th percentiles, respectively. The whiskers extend to the most extreme data points not considered outliers, and the outliers are plotted individually using the ``$+$'' symbol.

In the first Monte Carlo experiment we consider $\Nb=[\, 40\; 40\; 7\,]$. The results are depicted in Fig.~\ref{fig:res1}. As one can see, the best performance is given by R-F and B-F, i.e., the methods using (\ref{theta_est_m2}) which exploits the information coming from the dependence between the two measurements channels. The ``shift'' methods R-S and B-S provide only minor improvements in the performance. The reason can probably be explained as follows. The true phase shift $e^{-iM\theta_3}$ between the two measurement channels is fixed, while the compensation term depends on the variable $\omega_3$. Therefore, the weight is not correct unless $\omega_3=\theta_3$. Since our computations are only carried out for the variable $\omegab$ on a discrete grid, an exact phase cancellation almost never happens.

In the second Monte Carlo experiment we consider $\Nb=[\, 60\; 60\; 4\,]$, i.e., the two ULAs receiver modules have the number of antennas which coincides with the series-production automotive radar sensor considered in \cite[Table 1]{engels2017advances}. The results are depicted in Fig.~\ref{fig:res2}: the same observations as before can be drawn. In the third Monte Carlo experiment we consider $\Nb=[\, 70\; 70\; 3\,]$,  i.e., the two ULAs receiver modules have a very small number of antennas. The same observation as before still hold, see Fig.~\ref{fig:res3}. These experiments show that the information in the cross term of the multidimensional multivariate windowed periodogram (\ref{Phi_corr_windowed}) provide an estimator of $\thetab$, through (\ref{theta_est_m2}), which outperforms the one that uses only the auto-spectra. 

Finally, it is worth noting that in all the Monte Carlo experiments, the number of samples per pulse and the number of pulses are small making the estimation procedure computationally efficient, in particular, when computing the FFT. On average, with the grid sizes corresponding to Figs.~\ref{fig:res1}, \ref{fig:res2}, and \ref{fig:res3}, the time needed to run one trial is $0.16$, $0.19$, and $0.22$ s, respectively. The simulation environment is Matlab on a laptop with an Intel Core i5-4200U CPU and 3.6 GB of RAM.

\section{CONCLUSIONS AND FUTURE WORKS}\label{sec:concl} 

We have considered an integrated system of automotive modules for the target parameter estimation problem. Simulation results show that 
the cross term in the multidimensional multivariate windowed periodogram provides a remarkable improvement for estimating the target frequencies. On the other hand, the proposed periodogram gives an estimation error of the target frequencies characterized by some outliers. Possible extensions are to develop high-resolution spectral estimation techniques (cf. \cite{georgiou2001spectral}) as well as methods applicable for MIMO transmit arrays \cite{li2007mimo}. Another  possible extension is to consider the more general formulation in \cite[Section~II]{Georgiou-06}, instead of a finite superposition of sinusoids as in \eqref{y_meas_n}. The ULAs' measurements can be modeled as integrals
\begin{equation}\label{y_j_integral}
\begin{split}
y_1(\tb) & = \int_{\Tbb^3} a(\thetab) e^{ i \left( \innerprod{\thetab}{\tb} +\varphi(\thetab) \right)} \d\thetab + w_1(\tb) \\
y_2(\tb) & = \int_{\Tbb^3} a(\thetab) e^{ i \left( \innerprod{\thetab}{\tb} + M\theta_3 +\varphi(\thetab) \right)} \d\thetab + w_2(\tb),
\end{split}
\end{equation}
where, now the amplitude $a$ and the initial phase $\varphi$ are both functions of the frequency, and $\d\thetab:=\d\theta_1\d\theta_2\d\theta_3$.
Again we make the assumption that $\varphi(\thetab)$ follows a uniform distribution in $[-\pi,\pi]$ for any $\thetab\in \Tbb^3$, and $\varphi(\thetab)$ and $\varphi(\thetab')$ are independent if $\thetab\neq\thetab'$.\footnote{Notice that the assumption of independence may be questionable as one naturally wants to require the function $\varphi(\thetab)$ to be continuous in $\thetab$ in the mean-square sense. However, since we eventually deal with discrete spectra in the implementation, such an assumption is acceptable.}
The covariance lag has the expression
\begin{equation}\label{moment_eqn}
\Sigma_{\kb} = \int_{\Tbb^3} |a(\thetab)|^2 R \, e^{ i \innerprod{\thetab}{\kb}} \d\thetab + \tilde{\sigma}^2 \delta_{\kb,\zerob} I_2.
\end{equation}
Hence the multidimensional multivariate spectral density is $\Phi(\thetab)= (2\pi)^3 |a(\thetab)|^2 R + \tilde{\sigma}^2 I_2$ in which $(2\pi)^3$ is a normalization constant for the Lebesgue measure on $\Tbb^3$. Within this framework it would be possible to build a theory for the construction of the spectrum based on the moment equation \eqref{moment_eqn} whose left-hand side is replaced with the standard biased covariance estimates (see Section~\ref{sec:periodogram}) computed from the radar measurements. It is worth noting that some a priori information on the spectrum to be estimated can be used in the same spirit of the so called THREE-like estimators \cite{byrnes2000new,P-F-SIAM-REV,ZorzSep,KLapprox,BEL02}.
Once a spectrum is constructed, the frequency estimate can be obtained by \eqref{theta_est_m2}.

\balance
\bibliographystyle{IEEEtran}
\bibliography{references}

\end{document}